\documentstyle[editedvolume]{crckapb} 
\begin{opening}
\title{The Central Regions of the Galaxy and Galaxies: A Brief Summary}
\subtitle{}

\author{F. Combes}
\institute{Observatoire de Paris, DEMIRM\\
           61 Av. de l'Observatoire, F-75 014 Paris, France}
\end{opening}

\runningtitle{Summary Talk}

\begin{document}

\section{Introduction}
 This symposium has revealed considerable new progress
on nuclei of galaxies over the last three years. In this talk, I will 
try to point out the advances since the last
meeting on exactly the same subject held three years ago at Ringberg Castle
in Germany (see "The Nuclei of Normal Galaxies: Lessons from
the Galactic Center", ed. R. Genzel \& A. Harris, Kluwer).

It was particularly timely to organize this symposium on galactic nuclei,
now in Kyoto, for several important new points have come about:

\begin{itemize}

\item New steps in spatial resolution, due to the Hubble Space Telescope,
the adaptive optics now operational, and also radio millimeter interferometers,
that operate at 1mm with subarcsec beams,

\item Near-infrared rapid progress in cameras and spectrographic instruments

\item New results from satellites, either in X-rays, $\gamma$-rays, or
far-infrared (ISO)

\item Proper motions breakthrough (ESO, Keck, H$_2$O masers)

\end{itemize}          

It is worth emphasizing that our Galactic Center is a unique laboratory, since
we can probe a galactic nucleus with unrivaled spatial resolution
(1" = 0.04 pc; 1mas = 8 AU for VLBI). However, our position inside the 
galactic plane, where we see the nuclear disk almost edge-on, is not
favorable, and mutual comparisons between the Milky Way and external
galaxies are necessary for a better understanding of the morphology
and dynamical processes.

During all this captivating week, it has been obvious and clear 
why it is so interesting and rewarding to study galaxy nuclei: they are

\begin{itemize}

\item {\bf Fascinating} since tremendous amounts of energy are released,
either through AGN-type of activity, or through starbursts

\item {\bf Singular} In many cases, black holes are suspected to
exist in nuclei because of the kinematical measurements; the density 
profiles are cuspy, having or not a well-defined core,

\item {\bf Unique} The center of our Galaxy is the unique place of large-scale
intense magnetic phenomena. Also, for the dynamics of a galaxy, the center 
is the end-point of recursive dynamical phenomena, traced by nested
structures like bars or spirals. These structures take over to drive the
gas of the galaxy disk towards the unique center.

\end{itemize}          

Although our galaxy is neither an AGN nor a starburst (and does not fall
in the first category), we can take lessons applicable to external nuclei, 
since it has the two other characteristics.

\section{Existence of Massive Black-holes (MBH)}

May be the most important results that we have heard of this meeting
are those confirming through proper motions the large velocities in at least
two galactic nuclei (Milky Way and NGC 4258). In the Milky Way, the
proper motions are stellar, seen at near-infrared wavelengths, and have been 
obtained over 6 years at ESO-NTT ({\it Eckart, Genzel}) and only one year at 
Keck ({\it Ghez}). These establish that the high velocities measured in the
radial direction by spectroscopy also exist in the plane of the sky, 
and that the velocity dispersion is isotropic. The large velocities
found around the NGC 4258 nucleus through H$_2$O masers seen at VLBI resolution,
are also confirmed through proper motions to be compatible with rotation
({\it Nakai, Hernstein}). It appears now more and more certain that
a black hole of $\approx$ 2.5 10$^6$ M$_\odot$ exists in the center of
our Galaxy. Massive black-holes have stopped to play the "Arlesian", this 
actor of the play that never shows up, but that every people is talking
about. Of course MBH were suspected long before ({\it
Richstone}). Could compact stellar clusters still be an alternative
 solution to the MBH? No, certainly no for the Milky Way
and NGC 4258 ({\it Maoz}), although a direct proof will come from 
measuring relativistic velocities at a few Schwarzshild radii. Many other 
candidates (10-20) for MBH were presented ({\it Ford, Richstone}), but those
are not so tightly confirmed.

With the increased spatial resolution of the HST, 
the new discovery is the existence
of nuclear disks, almost omni-present at very small scale around these
active nuclei: examples are gaseous, NGC 4261, M81, M87 ({\it Ford}),
and stellar, NGC 4342, ... ({\it van den Bosch}), their sizes are of the
order of 100-300 pc.

In exploring the dynamics of nuclei, to deduce the possible existence of 
MBH, the main problem is the velocity anisotropy and existence of radial
orbits. A lot of progress has been reported in the analysis of distribution
functions: three-integral models are developped, 2D fields and fine details
in LOSVD (line of sight velocity distribution) are examined ({\it van der
Marel, van den Bosch}). Even in ellipticals and early-type galaxies,
links with bars and resonant rings can be made ({\it Emsellem}). 

For the future, now that MBH have been proven to exist, many exciting
questions should be considered: MBH demography ({\it Ho}), the MBH formation,
mergers of MBH, their influence on the dynamics of the galaxy, etc...
The relation between the mass of the MBH and the mass of the host-galaxy,
more exactly the mass of the spheroid host (the total mass for an elliptical,
the bulge mass for a spiral) has been advanced ({\it Richstone}), but is
questionable: the scatter is about a factor 100, and observational biases
are obvious. It is of prime importance to investigate such relations further
to understand the formation scenarios of MBH.

There has been also considerable progress in the theory: the main puzzle of
the confirmed MBH is that they are not violent AGN, in particular
the activity of our own galactic center is several orders of magnitude
lower than expected, from the gas present (available fuel). We have 
learned that there exists a magic solution, called ADAF (Advection Dominated
Accretion Flow), that satisfies all observational constraints ({\it Narayan}).
In brief, the gas is swallowed by the black hole before it has time to
radiate away its gravitationally acquired energy. The constraints on
thickness (H/R), or disk mass, or upper limit on perturbations of keplerian 
velocities are met ({\it Hernstein}), and may be the jet in 
NGC 4258 ? ({\it Falcke}). The spectrum of SgrA$^*$ has been successfully fit in
the frame of the stellar wind accretion model ({\it Melia}); this model has 
been shown to impeed accretion rate and thus solve the SgrA$^*$ puzzle 
({\it Coker}). The spectrum can be interpreted as optically thin synchrotron 
emission from mono-energetic electrons in a 
core/shell structure ({\it Duschl}), or may be three structures ({\it Falcke}).

\section{Starbursts and stellar populations}

Although our Galaxy is not a starburst,
there is evidence of recent star formation in the Galactic center, even though
high extinction makes all results uncertain. In the central few hundred
parsecs, a continuous low-level star formation has formed a flattened
stellar cluster, of about a million solar masses, which is not to be confused
with the center of the bulge ({\it Serabyn}). There is evidence of massive
star formation, for instance in the Quintuplet cluster ({\it Nagata}), 
that illuminates the pistol-shaped HII regions ({\it Figer}), and hot dust
has been mapped with ISOCAM. The clues given by the various stellar populations
on the age and chemical evolution of the Milky Way bulge are hard to read
({\it Rich}). Long-period miras (P$>$ 300 days, i.e. the 
young population) are tracing the bar; but there exist also metal-rich old 
stars, and there does not appear any correlation between abundances and 
kinematics. There has been certainly recycling of stars, so that the
bulge formed rapidly but not with a starburst. Many contradictions remain.
A question frequently asked: are the bulge and bar the same? There is no
concensus ({\it Ng}). The situation is quite different in 
M31, which appears not to have had the same history: there is a super
metal-rich population of globular clusters, enlighted by recent HST studies
({\it Jablonka}). 
Very enthusiastic talks have drawn our attention on the large number of
shells discovered in neutral ISM surveys ({\it Tsuboi, Hasegawa}). There
could be as many as 500-1000 supernovae or stellar winds driven shells,
attesting of a past starburst.

\smallskip

As for starbursts in external galaxies, a long-standing problem
has been to disentangle both types of activities, from an AGN or a nuclear
starburst. The predominance of one or the other is now studied with 
ISO SWS/LWS observations, through line ratios of highly ionized
species (AGN) like OIV, and low ionised (NEII) or "PAH" features at 7$\mu$
or molecular H$_2$ lines ({\it Lutz, Egami}). An interesting study was
reported about compared stellar populations in starbursts, seyferts and liners
({\it Joly}): the mean age of the populations appears to increase from
the starbursts (logically the youngest), to seyfert 2, then seyfert 1, than
liners. The existence of the two categories of seyferts would not be reducible 
to projection effects.

\section{Singularities}

The high spatial resolution of the HST has recently discovered that
some spheroids possess resolved cores, while others do not. The first
category corresponds to the large and luminous ellipticals, that are
also predominantly boxy and triaxial, while the second category is the 
prerogative of the smaller ones, predominantly disky and oblate.
 The underlying interpretation is that the smaller ones, more rotationally
supported, make the transition with spirals; throughout coalescence between 
smaller systems, the
biggest systems lose their rotation, shape, and binary black holes can 
create the cores in the merging ({\it Makino poster}).
However, we learned that in the Coma cluster, there was no effect of the
environment on these two classes, and even that the bimodality character
was not recovered ({\it de Jong}).

\section{Magnetic Phenomena}

The order of magnitude of the magnetic field in the galactic center
is $\approx$ 1 mG, a thousand times higher than in the disk ($\approx$
1$\mu$G). The magnetic pressure ($\propto$ B$^2$) is therefore considerable.
 However the gravitational force is dominant in molecular clouds
({\it Morris, Novak}). 

The galactic center is unique for the fantastic morphology of non-thermal
filaments (NTF); there are physical links between NTF, thermal filaments,
molecular clouds and HII regions. The orientation of the field is completely
different in the clouds and in the intercloud, and field lines reconnection at
cloud surfaces could be the mechanism to accelerate relativistic electrons to
produce the synchrotron emission of the NTF. The origin of this strong magnetic
field could be a proto-galactic primordial field, that is amplified through gas
radial inflow ({\it Morris}). In this domain, future progress is certainly 
expected in theory development, which has a wealth of observations
to compare with.

\section{Dynamics and Fuelling}

Bars and embedded nuclear bars have been known for a long time. In theory,
since gravity is scale-independent, and because these structures are
gravity-driven, there is no reason not to extrapolate to even smaller
structures, nested inside nuclear bars. We have heard of significant
progress in spatial resolution, either from millimeter interferometers
({\it Tacconi, Scoville}), or adaptive optics ({\it Rouan poster, Knapen}), and
it appears that the third-level bar structure might have been discovered
already in NGC 1068. We have now improved statistics in dynamical
morphologies: rings, bars, spirals, twin peaks, etc.. ({\it Kenney,
Sakamoto}). Since these nuclear structures are usually very rich in 
molecular gas, it is of prime importance to settle the question of
its excitation, or metallicity, which both are involved in the still
uncertain CO/H$_2$ ratio ({\it Sofue, Turner, Downes}). 

On the theoretical side, nested bars have been modelised, and much more 
details of the 
dynamical mechanisms are understood ({\it Friedli, Wada}), and models are 
confronted successfully to observations ({\it Garcia-Burillo}).
 We have heard that $m=1$ perturbations are ubiquitous ({\it Blitz}),
and new mechanisms based on non-linear wave coupling have
been proposed ({\it Masset}). More studies should be devoted to $m=1$
perturbation mechanisms in galaxies, that are much less well known than
the $m=2$ analogs. Future progress in this domain requires
more spatial resolution to probe the micro-structures embedded into
nuclear structures. This will clarify the connection between the dynamics
and the AGN activity, since only micro-structures are relevant. 
There may be different processus of accretion and duty cycles, according
to the absolute luminosity of the AGN ({\it Ulrich}).

\bigskip

I would like, on behalf of all participants, to warmly thank our chairman, Prof.
Yoshiaki Sofue, for the organisation of such a fruitful meeting.
We are also indebted to the co-chairmen, Reinhard Genzel and Mark Morris, and
very grateful to their co-organisor, Masato Tsuboi, for very efficient time
scheduling.

\end{document}